\documentclass[aps,pre,twocolumn,floats,showpacs,floatfix,superscriptaddress]{revtex4-1}
\usepackage{graphics,graphicx,dcolumn,bm,epic,eepic,float}
\usepackage{amssymb,amsmath,multirow,rotate,color,float,mciteplus}
\usepackage[latin1]{inputenc}
\usepackage{color}
\usepackage[dvips]{epsfig}

\usepackage{epstopdf}

\newcommand{\Arch }{\mathrm{Arch}}

\begin{document}

\title{Enhanced slip properties of lubricant-infused grooves}

 \author{Evgeny S. Asmolov}
\affiliation{A.N.~Frumkin Institute of Physical
Chemistry and Electrochemistry, Russian Academy of Sciences, 31
Leninsky Prospect, 119071 Moscow, Russia}
\affiliation{Institute of Mechanics, M.V.~Lomonosov Moscow State
University, 119991 Moscow, Russia}

 \author{Tatiana V. Nizkaya}
\affiliation{A.N.~Frumkin Institute of Physical
Chemistry and Electrochemistry, Russian Academy of Sciences, 31
Leninsky Prospect, 119071 Moscow, Russia}

\author{Olga I. Vinogradova}
\affiliation{A.N.~Frumkin Institute of Physical
Chemistry and Electrochemistry, Russian Academy of Sciences, 31
Leninsky Prospect, 119071 Moscow, Russia}
\affiliation{Department of Physics, M.V.~Lomonosov Moscow State University, 119991 Moscow, Russia}
\affiliation{DWI - Leibniz Institute for Interactive Materials, Forckenbeckstra\ss e 50, 52056 Aachen,
  Germany}

\date{\today}
\begin{abstract}

We ascertain the enhanced slip properties for a liquid flow over lubricant-infused unidirectional surfaces. This situation reflects many practical settings involving liquid flows past  superhydrophobic grooves filled with gas, or past grooves infused with another, immiscible, liquid of smaller or equal viscosity, i.e. where the ratio of lubricant and liquid viscosities, $\mu \leq 1$. To maximize the slippage, we consider deep grooves aligned with the flow. The (normalized by a texture period $L$) effective slip length, $b_{\mathrm{eff}}$, is found as an expansion to first order in protrusion angle $\theta$ about a solution for a flat liquid-lubricant interface.  Our results show a significant increase in $b_{\mathrm{eff}}$ with the area fraction of lubricant, $\phi$, and a strong decrease with $\mu$. By contrast, only little influence of $\theta$ on $b_{\mathrm{eff}}$ is observed. Convex meniscus slightly enhances, and concave - slightly reduces $b_{\mathrm{eff}}$ relative the case of a flat liquid-lubricant interface. The largest correction for $\theta$ is found when $\mu = 0$, it decreases with $\mu$, and disappears at $\mu = 1$. Finally, we show that lubricant-infused surfaces of small $\theta$ can be modeled as flat with patterns of local slip boundary conditions, and that the (scaled with $L$) local slip length at the liquid-lubricant interface is an universal function of $\phi$ and $\mu$ only.

\end{abstract}

\maketitle

\section{Introduction}
\label{sec_intro}

The design and fabrication of slippery lubricant-infused surfaces that provide a significant enhancement in drag reduction for a  flowing liquid  have received much attention in recent years. Enhanced slip properties of the solid texture are normally promoted by an infused lubricant~\cite{wong.ts:2011,nizkaya2014gas,solomon.br:2014,keiser.a:2017}.  The best known example of such a lubricant is probably a gas trapped by superhydrophobic (SH) textures, but it could also be another liquid, such as oil or water.  Such liquid-infused (LI) surfaces present their own scientific challenges, being potentially much more stable and robust compared to SH surfaces for use in various applications, including anti-biofouling~\cite{epstein.ak:2012} and ice-phobicity~\cite{kim.p:2012}.

Another active area of current research includes investigations of unidirectional textures~\cite{rothstein.jp:2010,vinogradova.oi:2012} since it is relevant to a variety of micro- and nanofluidics applications where such surfaces do not only dramatically reduce viscous resistance~\cite{ybert.c:2007,bocquet2007}, but could also be employed to separate particles~\cite{pimponi.d:2014,LabChip} or enhance their mixing rate~\cite{ou.j:2007,nizkaya.tv:2017}. Since the enhanced slip properties of surfaces are induced by the presence of an infused lubricant in a contact with liquid, an important ongoing challenge is of quantifying their  effective slip. \citet{Bazant08} have proven that regardless of the complexity of the texture there exists the `fast' direction of the greatest effective slip. For unidirectional surfaces it obviously corresponds to longitudinal alignment with the shear stress. There is a large literature describing attempts to provide a satisfactory theoretical model to describe slippage properties of longitudinal grooves. We mention below what we believe are the more relevant contributions.

A pioneering paper published by \citet{philip.jr:1972} applied idealized shear-free local boundary conditions at the lubricant sectors. This has led to a simple analytical equation, which relates an effective longitudinal slip length (normalized by texture period $L$), ${b}_{\rm eff}$, to  the area fraction $\phi$ of perfect slip stripes
\begin{equation}\label{eq:philip}
    {b}_{\rm eff}^P =\frac{1}{\pi} \ln \left[ \sec \left( \frac{\pi \phi}{2}\right) \right] .
\end{equation}
 During last decade several papers have tried to calculate corrections to this solution caused by a meniscus curvature.  \citet{sbragaglia.m:2007} have calculated the first-order correction to ${b}_{\rm eff}^P$ assuming that the curvature of the meniscus is small. \citet{crowdy.d:2010,crowdy.d:2016} has studied the same longitudinal problem in the limit of small $\phi$, but without restriction on the protrusion angle. \citet{schnitzer.o:2016} has extended these results to find asymptotic formulas valid at larger no-shear fractions.  There have also been numerical calculations, which are directly relevant~\cite{teo.sj:2010,ageev.ai:2018}. All these subsequent attempts at improvements of an earlier model~\cite{philip.jr:1972} have shed some light on the role of the meniscus curvature. We should recall, however, that non of these papers have tried to relax the assumption of  shear-free liquid-gas interface. In other words, the effect of gas or of another lubricant confined in the grooves has been fully ignored.

The body of theoretical and experimental work investigating flows past more general LI surfaces is much less than that for SH surfaces, although there is a growing literature in this area. \citet{ng.co:2010} have investigated lubricant-infused grooves and shown that even small lubricant viscosity may affect the effective slip length predicted by \citet{philip.jr:1972}. We remark that these authors have not included a meniscus curvature into consideration. To account for a dissipation within the lubricant several groups suggested to replace the
two-phase approach with a single-phase problem with  partial slip boundary condition
imposed at the flat lubricant areas~\cite{cottin.c:2004,vinogradova.oi:2010,asmolov_etal:2013,asmolov:2013b}.  \citet{belyaev.av:2010a} have derived an expression for a longitudinal effective slip length of surfaces decorated by partially slipping stripes
\begin{equation}\label{eq:belyaev}
  {b}_{\mathrm{eff}}^{BV}\simeq\dfrac{1}{\pi}\dfrac{\ln\left[\sec\left(\frac{\pi\phi}{2}\right)\right]}{1+\dfrac{1}{\pi {b}_c}\ln\left[\sec\left(\frac{\pi\phi}{2}\right)+\tan\left(\frac{\pi\phi}{2}\right)\right]},
\end{equation}
where ${b}_c$ is a constant local slip length (scaled with $L$). In the limit ${b}_c \to \infty$, which is equivalent to shear-free boundary conditions, Eq.(\ref{eq:belyaev}) reduces to the solution by \citet{philip.jr:1972},  but it predicts smaller ${b}_\mathrm{eff}$ when ${b}_c$ is finite. \citet{ng.co:2011} have assumed that the curved meniscus interface has a constant partial slip length ${b}_c$, and then calculated the effective slip semi-analytically. Neither papers attempted to properly connect ${b}_c$ with the viscous dissipation in the infused lubricant, but since ${b}_c \propto \mu^{-1}$, where $\mu $ is a ratio of the dynamic viscosities of
the lubricant and the liquid, we consider they shed some important light on the role of a lubricant viscosity.

 Several theoretical papers have been concerned with the infused lubricant effect on the local slip length.  \citet{hocking.lm:1976} has concluded that the local slip length of lubricant-infused irregularities is proportional to their depth if shallow and to their spacing if deep. \citet{shoenecker.c:2014,shoenecker.c:2013} have argued that the distribution of a local slip length across the lubricant-fluid interface is
non-uniform. \citet{nizkaya2014gas} have elucidated  a mechanism which transplants the flow in the lubricant to a local slip boundary condition at the fluid-lubricant interface. This study has concluded that the   non-uniform longitudinal local slip length of a shallow texture is defined by the viscosity contrast and local thickness of a thin lubricating films, similarly to infinite systems~\cite{vinogradova.oi:1995a,miksis.mj:1994}. By contrast, a (divided by $L$) non-uniform local slip length at a lubricant interface of a deep texture, ${b}$, can be expressed as~\cite{nizkaya2014gas}
\begin{equation}
b \simeq \dfrac{\phi \beta(y/\phi )}{\mu}.  \label{bgas}
\end{equation}%
Here  $\beta$ denotes the non-uniform slip coefficient. These papers appear to have made an important contribution to the subject, but again, no attempt has been made to include the meniscus curvature in the analysis.

Thus, a quantitative understanding of liquid friction past LI (and even SH) grooves remains challenging. Although it is now clear that both dissipation in the lubricant and the curvature of the liquid-lubricant interface may simultaneously affect lubricating properties of the surfaces, the investigation of these two effects in the current literature is decoupled. Researchers studying the role of meniscus appear to fully ignore the viscous dissipation, while others investigate the viscous dissipation by excluding the meniscus from the analysis. We are unaware of previous work that has addressed the question of effective and local slip calculations in the situation when both the lubricant viscosity and meniscus curvature may be important. The only exception is probably a very recent study ~\cite{crowdy.d:2017b}, where integral expression for the correction to Eq.(\ref{eq:philip}) due to weak meniscus curvature has been proposed. However, this has been done for $\mu \ll 1$, which is the case of SH surfaces only, and the viscosity ratios in real experiments and applications involving LI surfaces can be much larger~\cite{solomon.br:2014,keiser.a:2017}.

In this paper we offer theoretical insights on the general situation, where both weak meniscus curvature and the viscosity contrast between liquid and lubricant phases are taken into account. We consider shear flow past unidirectional periodic texture, varying on scales smaller than the
channel thickness. The geometry of deep rectangular grooves~\cite{dubov.al:2017} and  their longitudinal alignment\cite{Bazant08} with the shear stress have been chosen to maximize the effective slip length of a flat interface. Our focus here is on a situation, when a lubricant is of smaller viscosity than a liquid, which is expected to induce enhanced slip properties~\cite{solomon.br:2014,keiser.a:2017}. Another special topic here is LI surfaces, where a lubricant and a liquid are of the same viscosity, and we compare their friction properties with predicted for a situation, when the liquid follows the topological variations of the surface~\cite{wang2003}. Our theory is based on a perturbation approach~\cite{sbragaglia.m:2007}, and we construct the first-order corrections due to a meniscus curvature to a longitudinal effective slip length of a flat unidirectional lubricant-infused surface.

The paper is organized as follows. In Sec.~\ref{geqs} we formulate the
governing equations and boundary conditions for two-phase and single-phase problems of calculation of velocity fields. The details of calculations of the effective slip length are given in Sec.~\ref{calc}. Sec.~\ref{res} contain results of our numerical calculations. We conclude in Sec.~\ref{concl} with a discussion of our main results and their possible extensions. Appendix~\ref{ap:A} contains a derivation of boundary conditions at a curved liquid-lubricant interface. In Appendix~\ref{ap:B} we derive an analytical expression, which describes a correction to the effective slip length in the shear-free case.

\begin{figure}
\centering
\includegraphics[width=0.9\columnwidth]{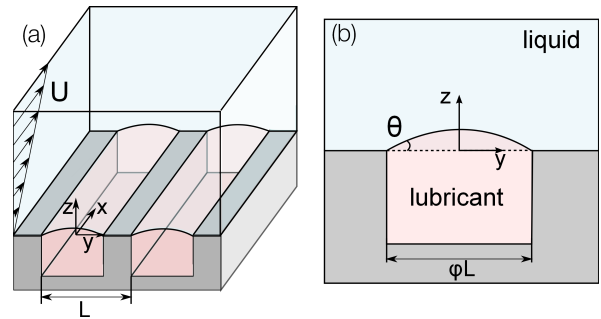}
\caption{(a) Longitudinal semi-infinite shear flow of
liquid of viscosity $\mu^*$ over a periodic array of rectangular grooves containing a lubricant of viscosity $\mu_l^*$. (b) Single period window for the grooves. }
\label{fig_sketch}
\end{figure}

\section{Governing equations}
\label{geqs}
%\section{Model}
%\label{model}

Fig.~\ref{fig_sketch} shows a schematic of our system.
We consider a longitudinal flow of a liquid of viscosity $\mu^*$ and density $\rho^*$ past an unidirectional texture infused with a lubricant of viscosity $\mu_l^*$, and assume the viscosity ratio
\begin{equation}\label{viscosity}
    \mu \equiv \frac{\mu^*_l}{\mu^*} \leq 1.
\end{equation}
The period of the texture is $L$, so that the meniscus occupies width $\phi L$. In our model the contact line is pinned to the sharp edge of the rectangular grooves, which are chosen to be deep to maximize the slippage. We treat the case of a thick channel (or of a single interface),
so that the liquid velocity profile sufficiently far from the meniscus may be considered as a linear shear flow of a rate $G$.

 We use $L$ as a reference length scale, so that all variable are dimensionless and fluid velocities are scaled by $GL$. We focus on a flow of low $\mathrm{Re}=\rho ^{\ast
}GL^{2}/\mu ^{\ast }$, where  $\mu^*/\rho^*$ is the kinematic viscosity.

Since the flow is periodic our  attention is restricted to the single period
window. We use Cartesian coordinate
system ($x$, $y$, $z$) with the $x$-axis parallel to the groove. The cross-plane coordinates are $y$ and $z$. We locate $y=0$ at the midplane of the groove and define the flat solid-liquid interface at
\begin{equation}\label{y_solid}
 \phi /2<\left\vert y\right\vert <1/2, \, z = 0
\end{equation}

We denote the protrusion angle with respect to the horizontal as $\theta $. It is defined as positive, when a lubricant protrudes into the liquid (convex meniscus), and as negative when liquid protrudes inside the groove (concave meniscus). We assume that the
meniscus is only weakly deformed from the flat state, so that $\left\vert \theta \right\vert \ll 1$. Therefore, the curved meniscus interface is expressed as
\begin{equation}\label{y_meniscus}
    \left\vert y\right\vert \leq \phi /2, \, z=\theta \eta,
\end{equation}
where dimensionless function $\eta \left( y\right)$ describes the shape of the meniscus, which represents the arc of the circle of the dimensionless radius $R=\phi /\left( 2\theta
\right) \gg 1$: $\left( \theta \eta +R\cos \theta \right) ^{2}+y^{2}=R^{2}$. For small $\theta$ we then easily obtain
\begin{equation}
\eta =\phi /4-y^{2}/\phi,  \label{eta}
\end{equation}
which can be substituted into Eq.(\ref{y_meniscus}).

%\section{Governing equations}

The problem is homogeneous in
$x$ direction ($\partial_x = 0$), In this case the velocity field $u(y,z)$ of both phases can be determined by solving the Laplace
equation
\begin{equation}
\Delta u=0.  \label{se}
\end{equation}
We stress that since Eq.(\ref{se}) does not contain a pressure term, its solution remains valid for any capillary number, $Ca$.

At the solid-liquid interface, which location is defined by Eq.(\ref{y_solid}), we apply the no-slip boundary conditions for the liquid velocity field, $u=0$.
%\begin{equation}
%u=0.  \label{ns}
%\end{equation}
The lubricant velocity at the side walls also satisfies the no-slip conditions.

At a curved interface, defined by Eqs.(\ref{y_meniscus}) and (\ref{eta}), we impose the boundary conditions of the continuity for the
velocity and tangential stress,
\begin{equation}
u=u_{l},  \quad
\left( \mathbf{n\cdot \nabla }\right) u=\mu \left( \mathbf{n\cdot \nabla }%
\right) u_{l},  \label{shstr}
\end{equation}%
where $u_{l}$ is the velocity of an infused lubricant and
\begin{equation}
\mathbf{n}\simeq \left( 0,-\theta \partial_y \eta,1\right),  \label{ntau}
\end{equation}%
is the unit normal vector, and $\partial_y \eta = 2 y / \phi$ is an outward normal derivative on the curved meniscus.

\begin{figure}[tbp]
\centering
\includegraphics[width=0.9\columnwidth]{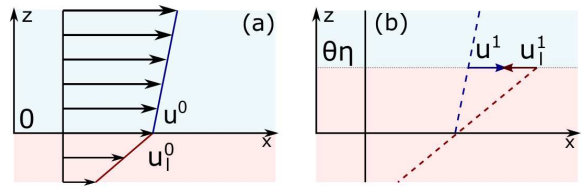}
\caption{(a) Viscous flow near a flat liquid-lubricant interface located at $z=0$. The leading-order velocities $u^0$ and $u^0_l$ at this boundary are equal; (b) Schematic illustration of $u^1$ and $u^1_l$ at the lubricant-meniscus interface, $z=\theta \eta$.
}
\label{fig_ap1}
\end{figure}

For a small protrusion angle, $\theta \ll 1$, the velocity can be expanded about $u^0$ and, to first order in $\theta$:
\begin{equation}
u \simeq U+u^{0}+\theta u^{1},
\label{exp}
\end{equation}%
where $U=z\ $ is the velocity of an undisturbed linear shear flow, $u^{0}$ is the zero-order solution for a flat liquid-lubricant interface (shown schematically in Fig.~\ref{fig_ap1}(a)), and $\theta u^1$ is the first-order correction due to a meniscus curvature. Both $u^{0}$ and $u^{1}$ vanish as $z \to \infty$, i.e. in the bulk liquid.

Let us first formulate boundary conditions, which should be imposed to obtain the zero-order solution for a velocity field (see Fig.~\ref{fig_ap1}(a)). In this case $\mathbf{n}^{0}=\left( 0,0,1\right)$, so the boundary conditions (\ref{shstr}) for two-phase problem can be written as
\begin{equation}
u^{0}=u^{0}_l,  \label{vt0}
\end{equation}%
\begin{equation}
\partial _{z}u^{0}-\mu \partial _{z}u^{0}_l=-\partial _{z}U=-1.  \label{sh0}
\end{equation}%
Note that when $\mu =1$, the shear rates at the liquid-lubricant interface in both phases are equal, $\partial _{z}(U+u^{0})= \partial _{z}u^{0}_l$. This means that the problem is fully identical to that of a single-phase flow over grooved surface considered earlier by~\citet{wang2003}.
When $\mu = 0$, we have $\partial _{z}u^{0}=-1$, i.e. the  liquid-lubricant interface is shear-free, and we recover the problem of  \citet{philip.jr:1972}.

We now formulate the boundary conditions, which have to be applied to calculate $u^1$. For the solid-liquid interface we should naturally impose the condition $u^{1}=0$.
For the meniscus (schematically shown in Fig.~\ref{fig_ap1}(b)) we obtain the following boundary conditions (see  Appendix~\ref{ap:A} for a derivation)
\begin{equation}
u^{1} = u^{1}_l+\left( 1/\mu -1\right) \eta (\partial _{z}u^{0}+1),  \label{sb5}
\end{equation}%
\begin{equation}
\partial _{z}u^{1}-\mu \partial _{z}u^{1}_l =  \left( 1-\mu \right) \partial
_{y}\left( \eta \partial _{y}u^{0}\right) .  \label{sb1}
\end{equation}%
Thus, the boundary conditions to the zero- and first-order solutions formulated for $z=0$ are similar (cf. Eqs.(\ref{vt0}), (\ref{sh0}) and (\ref{sb5}), (\ref{sb1})). The only difference is the velocity jump in Eq.(\ref{sb5}) and the shear rates inducing flows (right-hand sides in (\ref{sh0}) and (\ref{sb1})).

Alternatively, we can  replace the two-phase approach with a single-phase problem with spatially
dependent partial slip boundary condition, and express the lubricant shear rate in terms of a local slip length~\cite{nizkaya2014gas}
\begin{equation}
\partial _{z}u^{0}_l=\frac{u^{0}_l}{\phi \beta}.  \label{sh2}
\end{equation}%
The conditions (\ref{vt0}) and (\ref{sh0}) can then be reduced to
\begin{equation}
\partial _{z}u^{0}-\frac{\mu u^{0}}{\phi \beta}=-1.  \label{sh1}
\end{equation}
%Thus we have two- and single-phase problems with the boundary conditions (\ref{vt0}), (\ref{sh0}) and (\ref{sh1}), respectively, where the flow is %induced by a given shear rate $\partial _{z}U=1$.

For the first-order problem Eqs.(\ref{sb5}) and (\ref{sb1}) can be also formulated in terms of the local slip length,
\begin{equation}
u^{1} = u^{1}_l+\left( 1-\mu \right)
\frac{\eta u^{0}}{\phi \beta}, \label{sb5_b}
\end{equation}%
\begin{equation}
\partial _{z}u^{1}-\frac{\mu u^{1}}{\phi \beta}=\left( 1-\mu \right) %
\left[ \partial _{y}\left( \eta \partial _{y}u^{0}\right) -\frac{\mu \eta
u^{0}}{\phi ^{2}\beta^{2}}\right] .  \label{sb6}
\end{equation}
Here to transform Eq.(\ref{sb1}) to Eq.(\ref{sb6}) we made an assumption that the disturbances of the velocity and of the shear rate in the lubricant are related as $\phi \beta\partial
_{z}u^{1}_l=u^{1}_l$, similarly to Eq.(\ref{sh2}). Eq.(\ref{sb6}) allow us to replace again the two-phase problem by the single-phase one using the same profile of the local slip length as for the flat interface.

We note that when
$ \mu = 0$, both conditions (\ref{sb1}) and (\ref{sb6})  reduce to derived by~\citet{sbragaglia.m:2007} in the shear-free limit
\begin{equation}
 \partial _{z}u^{1}= \partial _{y}\left(
\eta \partial _{y} u^{P}\right),  \label{f_u}
\end{equation}%
We also stress that when  $\mu = 1$ the terms in the right-hand sides of Eqs.(\ref{sb5}) and (\ref{sb1}), which induce the disturbance flow, vanish, so that
we get $u^{1} = u^{1}_l=0$. This implies that independently on the shape of a meniscus the flow remains fully identical to a single-phase flow over grooved surfaces~\cite{wang2003}.  This case can also be described in terms of a local slip length~\cite{nizkaya2014gas}.

%\subsection{Asymptotic limits}

\section{Calculations of effective slip lengths}
\label{calc}

We calculate the dimensionless effective slip length at $z=0$ as an expansion,%
\begin{equation}
b_{\mathrm{eff}}\simeq b^{0}_{\mathrm{eff}}+\theta b^{1}_{\mathrm{eff}}.  \label{beff}
\end{equation}
 Here $b^{0}_{\mathrm{eff}}$ is the zero-order solution for a flat liquid-lubricant interface, and $\theta b^{1}_{\mathrm{eff}}$ is the first-order correction due to a meniscus curvature, which is related to the liquid velocity at the liquid-lubricant interface as
\begin{equation}
b_{\mathrm{eff}}^{1}=\int_{-\phi /2}^{\phi /2}u^{1}\left( y,0\right) dy.
\label{b1ef}
\end{equation}

To find $u^1$ we construct the solution of Stokes equations in terms of Fourier series. Since the velocity is an even function of $y$, a general solution of the Laplace equation decaying at infinity has the form
\begin{equation}
u^{1}=\frac{c_{0}}{2}+\sum_{n=1}^{\infty }c_{n}\cos \left( k_{n}y\right)
\exp \left( -k_{n}z\right) ,
\label{four}
\end{equation}%
with $k_{n}=2\pi n$.

To solve a two-phase problem we should similarly expand the solution for a lubricant flow within grooves in Fourier series,  but with $k_{m}=2\pi  \left(
2m-1\right) /\phi$:%
\begin{equation}
u^{1}_l=\sum_{m=1}^{\infty }c_{m}^{l}\exp \left( k_{m}^{l}z\right) \cos
\left( k_{m}^{l}y\right) .
\label{exp_l}
\end{equation}%
Since $\cos\left( k_{m}^{l}\phi /2\right)=0$ for any $k_{m}^{l}$, the expansion (\ref{exp_l}) enable us to satisfy the no-slip condition at the side
walls automatically. %Normal derivative of the lubricant velocity is then: $\partial_z u^1_l=\displaystyle\sum_{m=1}^{\infty } k_{m}^{l}c_{m}^{l}\exp \left( k_{m}^{l}z\right) \cos\left( k_{m}^{l}y\right)$. This yields the following
The vector of normal derivatives $\partial_z u^{1}_l(y_j)$ at collocation nodes $y_j$ can be connected to the vector of velocities $u^{1}_l(y_j)$ via the Dirichlet-to-Neumann matrix derived in ~\cite{nizkaya2014gas}. Then the coefficients $c_{n}$ and $c_{m}^{l}$ are found by applying boundary conditions given by Eqs.(\ref{sb5}) and (\ref{sb1}).

To calculate the Fourier coefficients $c_n$ within the single-phase approach  we apply a collocation method on a uniform grid spanning $|y|<1/2$, and by satisfying boundary conditions Eq.(\ref{sb6}) pointwise. In these calculations we use
\begin{equation}
\beta =0.4-1.29(y/\phi )^{2}-1.24(y/\phi )^{4},
 \label{beta}
 \end{equation}
which is obtained by fitting the local slip coefficient of deep grooves found before~\cite{nizkaya2014gas}.
%$$\beta =[1/4-(y/\phi )^2][1.24(y/\phi )^2+1.6].$$

%\textbf{Operator method relating $c_{n}$ and $c_{m}^{l}$ and collocations}

For the shear-free limit, we use the analytical solution for the liquid velocity at a flat interface found by~\citet{philip.jr:1972}
\begin{equation}
u^{P}(y,0)=\frac{1}{\pi }\Arch \left[ \frac{\cos \left( \pi y\right) }{\cos
\left( \pi \phi /2\right) }\right] ,  \label{u0}
\end{equation}%
so that
\begin{equation}
\partial _{y} u^{P}=-\frac{\sin \left( \pi y\right) }{\sqrt{\cos ^{2}\left( \pi
y\right) -\cos ^{2}\left( \pi \phi /2\right) }}.  \label{u0p}
\end{equation}%

In this case $b^{1}_{\mathrm{eff}}$ may be determined from~\cite{sbragaglia.m:2007,crowdy.d:2017b}:
\begin{eqnarray}
b^{1}_{\mathrm{eff}} &=&\int_{-\phi /2}^{\phi /2}\frac{\left[ 1-\cos \left( 2\pi
y\right) \right] \eta }{\cos \left( 2\pi y\right) -\cos
\left( \pi \phi \right) }dy \notag \\
&=&2\int_{0}^{\phi /2}\eta (\partial_y u^{P})^{2} dy.  \label{sb4}
\end{eqnarray}%

We should like to mention that \citet{teo.sj:2010} have calculated the effective slip length for a shear-free case numerically and have shown that the linearized approximation, Eq.(\ref{beff}), is very accurate when $|\theta |\leq \pi/6$. More precisely, with these values of $\theta$ its deviation from exact numerical results is below 5\%. By this reason below we vary $\theta$ in this interval.

\section{Results and discussion}
\label{res}

\begin{figure}[h!]
\centering
\includegraphics[width=\columnwidth]{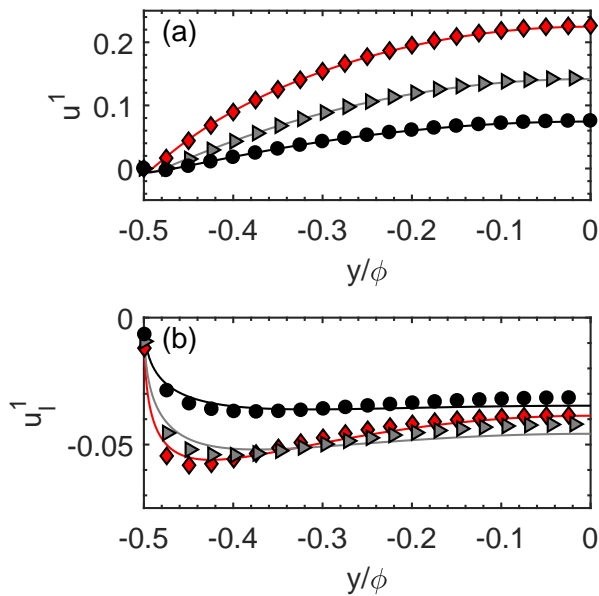}
\caption{Velocities, $u^1$ (a) and $u^1_l$ (b), vs. $y/\phi$ computed within the two-phase approach at fixed $\mu =0.2$ and $\phi =0.5$ (circles), 0.75 (triangles), 0.9 (diamonds). Solid curves plot the results obtained within the single-phase model.}
\label{fig2}
\end{figure}

%\textbf{Is $u^{1SP}$ quadratical in $u^0_{max}$ (B3) or linear (fig.5b)?}

We begin by studying velocities $u^1$ and $u^1_l$ computed by using two- and single-phase approaches. Fig.~\ref{fig2} shows $u^1$ and $u^1_l$ as a function of $y/\phi$ calculated at fixed $\mu=0.2$, which is the case of a typical oil-water interface, and several $\phi $. We see that $u^1$ is always positive and increases with $\phi$. In contrast, $u^1_l$ is negative. This result reflects the velocity jump in Eq.(\ref{sb5}) (see also Fig.~\ref{fig_ap1}(b)). Remarkably, the curves for $u^1$ obtained by using the local slip length concept practically coincides with the exact solutions of the two-phase problem as seen in Fig.~\ref{fig2}(a), but note that there is some small discrepancy in $u^1_l$ obtained with these two approaches (see Fig.~\ref{fig2}(b)). However, these results generally suggest that the weakly curved liquid-lubricant interface can be successfully  modeled as a pattern of a local slip boundary condition imposed at $z=0$.

\begin{figure}[h!]
\centering
\includegraphics[width=\columnwidth]{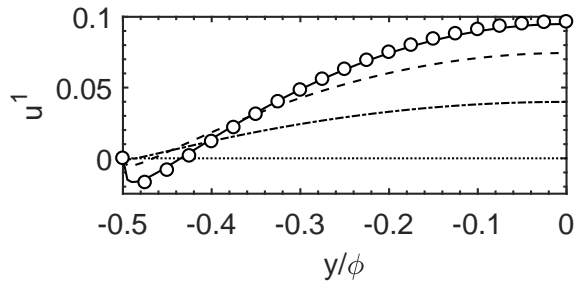}
\caption{Velocity $u^1$ calculated at $\phi =0.5$. Circles show results for $\mu =0$. Solid, dashed, dash-dotted, and dotted curves plot results for $\mu = 0.02$, 0.2, 0.5, and 1.}
\label{fig2c}
\end{figure}

We now fix $\phi = 0.5$ and explore how the viscosity contrast influences $u^1$.  Fig.~\ref{fig2c} presents the results obtained with several typical viscosity contrasts, varying from $\mu=0$ to $\mu = 1$. We conclude that at a very small  viscosity contrast, $\mu =0.02$, which is the situation of a water-air interface, $u^1$ is very close to that for $\mu = 0$, where it is largest. This correction decreases with $\mu $ and vanishes when $\mu =1$. We note, that near the edge of the grooves $u^1$ becomes negative, which implies that in this region $\theta u^1$ is negative for a convex meniscus and positive for a concave one.  However, since $u^1$ is positive for a major portion of the liquid-lubricant interface, it is obvious that $b_{\mathrm{eff}}^{1}$ given by Eq.(\ref{b1ef}) should always be positive. Therefore, positive $\theta$ do lead to a positive first-order correction to $b_{\mathrm{eff}}^0$, i.e. enhance the effective slip, but negative $\theta$ could only reduce its value.

\begin{figure}[h!]
\centering
\includegraphics[width=\columnwidth]{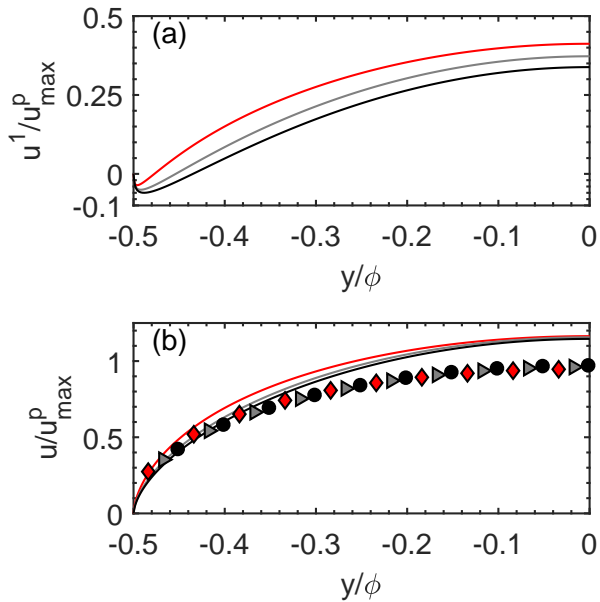}
\caption{(a) Normalized velocities, $u^1/u_{\max }^{P}$, calculated for $\mu=0.02$. Solid curves from top to bottom plot results for $\phi=0.9$, $0.75$ and $0.5$. (b) Corresponding normalized velocities, $u/u_{\max }^{P}$, computed with $\theta=\pi/6$.  Symbols show $u^0/u_{\max }^{P}$. }
\label{fig_deltabeff}
\end{figure}

Since velocity $u^1$ calculated at the water-air interface (see Fig.~\ref{fig2c}) is very close to that found in a shear-free case, $\mu = 0$, it is instructive to compare the velocity profiles obtained for these two cases in more detail. Velocity $u^1$ takes its maximum at $y=0$, and for the case of $\mu = 0$ and $\theta = 0$ its value can be easily obtained from Eq.~(\ref{u0})
\begin{equation}
u_{\max }^{P}=u^{P}(0,0)=\frac{1}{\pi }\Arch \left[ \sec \left( \frac{\pi \phi }{2}\right) \right] .
\label{upm}
\end{equation}
We now compute $u^0$, $u^1$, and $u$ for several $\phi$ and normalize them by $u_{\max }^{P}$. The results are presented in Fig.~\ref{fig_deltabeff}. We remark and stress that $u^0/u_{\max }^{P}$ obtained for chosen values of $\phi$ nearly coincide, and in fact they are well described by the elliptic velocity profile derived by \citet{philip.jr:1972} for small $\phi$. We return to the importance of this finding later, by discussing the effective slip length. The normalized velocity, $u^1/u_{\max }^{P}$, grows with $\phi$, and we also observe that the region of a negative $u^1/u_{\max }^{P}$ decreases with $\phi$. Also included are liquid velocity profiles $u=u^{0}+\theta u^{1}$ calculated for $\theta=\pi/6$. We conclude that they weakly depend on $\phi$ and that the effect of $\theta$ on $u$ is well pronounced. In this example, which corresponds to a convex meniscus, velocities $u$ are well above $u^0$. For a concave meniscus, they, of course, become smaller than $u_0$.

\begin{figure}[h!]
\centering
\includegraphics[width=\columnwidth]{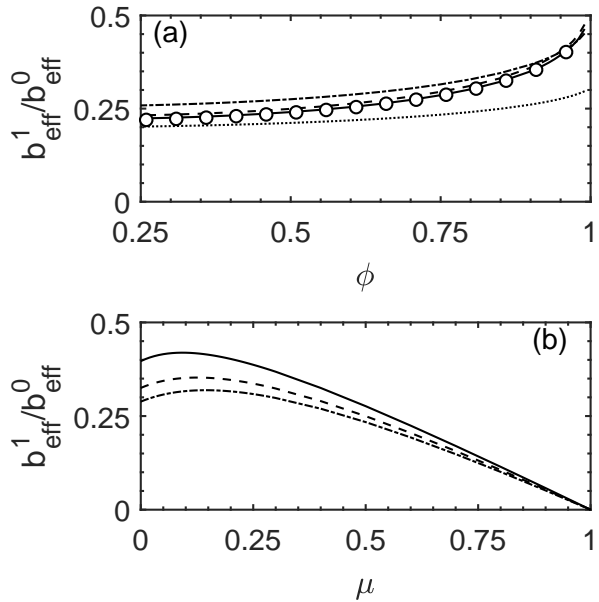}
\caption{(a) The ratio $b_{\mathrm{eff}}^{1}/b_{\mathrm{eff}}^{0}$ as a function of $\phi$. Solid, dashed, dash-dotted, and dotted curves correspond to $\mu=0$, $0.02$, $0.2$, and $0.5$. Symbols show calculations from Eq.(\ref{sb4}). (b) The same, but as a function of $\mu$. Solid, dashed, dash-dotted curves correspond to $\phi=0.9$, $0.75$, and $0.5$. }
\label{fig_beff}
\end{figure}

In Fig.~\ref{fig_beff}(a) we plot the ratio $b_{\mathrm{eff}}^{1}/b_{\mathrm{eff}}^{0}$ as function of $\phi$. It is seen that for all $\mu$ this ratio increases with $\phi$. In other words, the role of meniscus curvature is more pronounced at larger lubricant area. When $\mu=0$, results of our calculations coincide with obtained from Eq.(\ref{sb4}), confirming the validity of our approach. In Appendix \ref{ap:B} by using the fact that profiles of $u^0$ are elliptic for all realistic values of $\phi$ (Fig.~\ref{fig_deltabeff}(b)) we derive a simple analytical formula, Eq.(\ref{ab3}), describing accurately $b^{1}_{\mathrm{eff}}$ in this case up to $\phi\leq 0.9$. Eq.(\ref{ab3}), in particular, predicts that $b_{\mathrm{eff}}^{1}/b_{\mathrm{eff}}^{P} \simeq 2/3\pi$ at relatively small $\phi$, which is confirmed in Fig.~\ref{fig_beff}(a). At a small viscosity contrast $b_{\mathrm{eff}}^{1}/b_{\mathrm{eff}}^{0}$ grows relative to the case of $\mu=0$ except the case $1-\phi \ll 1$. On increasing $\mu$ further $b_{\mathrm{eff}}^{1}/b_{\mathrm{eff}}^{0}$ decreases and for all $\phi$ becomes smaller than expected at $\mu=0$. To examine this effect in more detail in Fig.~\ref{fig_beff}(b) we plot $b_{\mathrm{eff}}^{1}/b_{\mathrm{eff}}^{0}$ as a function of $\mu$. We see that all curves have their maxima at relatively small $\mu$. If we reduce $\phi$, the maximum at  $b_{\mathrm{eff}}^{1}/b_{\mathrm{eff}}^{0}$ is less pronounced and shifted towards larger $\mu$. When $\mu \geq 0.5$, $b_{\mathrm{eff}}^{1}/b_{\mathrm{eff}}^{0}$ decays linearly with $1 - \mu$ and the slope of these lines slightly depends on $\phi$. At $\mu = 1$ all curves vanish, which implies that the first-order correction to the zero-order effective slip length disappears, so that in this limit the meniscus does not affect the flow compared to the case of $\theta=0$.

\begin{figure}[h!]
\centering
\includegraphics[width=\columnwidth]{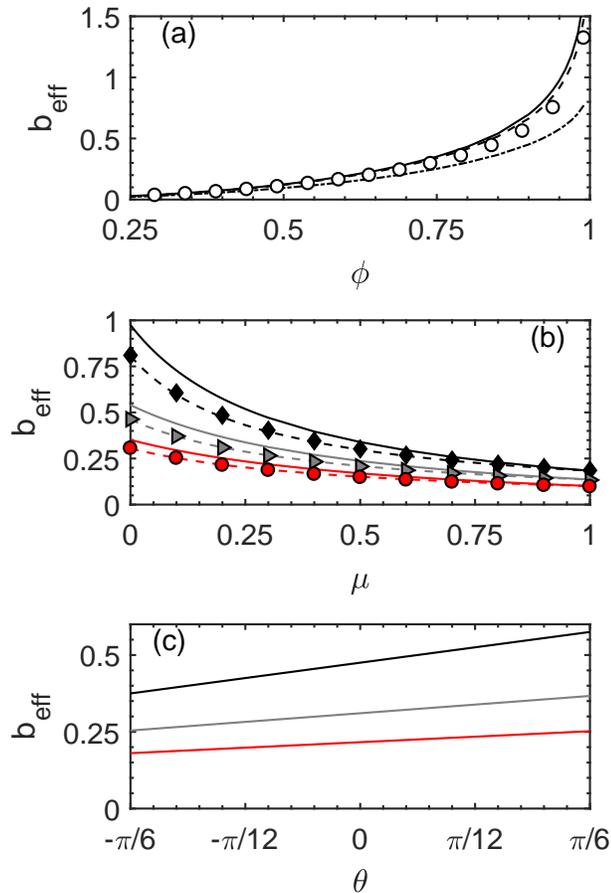}
\caption{(a) Effective slip length vs. lubricant fraction $\phi$. Calculations are made for $\theta=\pi/6$. Solid, dashed, dashed-dotted curves correspond to  $\mu=0$, $0.02$ and $0.2$. Symbols show $b_{\mathrm{eff}}^P$ given by Eq.(\ref{eq:philip}). (b)  The same as a function of $\mu$. From top to bottom $\phi=0.9$, $0.75$ and $0.5$. Solid and dashed curves plot results obtained for $\theta=\pi/6$ and $0$. Symbols denote $b_{\mathrm{eff}}^{BV}$ calculated from Eq.(\ref{eq:belyaev}) using $b_c=0.323\phi /\mu$. (c) The same as a function of $\theta$ calculated with $\mu=0.2$. From top to bottom $\phi=0.9$, $0.75$ and $0.5$.}
\label{fig_beff2}
\end{figure}

Finally, we calculate the effective slip length, $b_{\mathrm{eff}}$, and the results are illustrated in Fig.~\ref{fig_beff2}. Fig.~\ref{fig_beff2}(a) shows $b_{\mathrm{eff}}$ as a function of $\phi$. The results are obtained at fixed $\theta = \pi/6$ and several $\mu$. The effective slip length, $b_{\mathrm{eff}}^P$ calculated from Eq.(\ref{eq:philip}) is also shown. For typical SH surfaces with small curvature of a convex meniscus, $\mu=0$ and $0.02$, the effective slip length is seen to be slightly larger than $b_{\mathrm{eff}}^P$. However, when $\mu = 0.2$ it is much smaller compared to $b_{\mathrm{eff}}^P$. Note that in all cases $b_{\mathrm{eff}}$ strongly increases with $\phi$. It also significantly (monotonically) decreases with $\mu$, as it can be seen in Fig.~\ref{fig_beff2}(b), where $b_{\mathrm{eff}}$ calculated with $\theta=0$ and $\theta=\pi/6$ are plotted as a function of viscosity contrast for several fixed $\phi$. This plot also demonstrates that the effect of $\theta$ on $b_{\mathrm{eff}}$ is largest when $\mu=0$ and lubricant area, $\phi$, is large, but it reduces strongly with $\mu$ and with a decrease in $\phi$. It is well seen that it is getting extremely small when $\mu$ is above $0.5$, and it fully disappears at $\mu=1$. We also stress that the first-order correction to $b_{\mathrm{eff}}^0$ does not seem to be significant enough to be taken into account at any $\mu$ when $\phi=0.5$ (and, naturally, smaller). This implies that in many practical situations one can use Eq.(\ref{eq:belyaev}) to very accurately predict $b_{\mathrm{eff}}$ of lubricant-infused surfaces. Indeed, theoretical curves calculated from Eq.(\ref{eq:belyaev}) using $b_c=0.323\phi /\mu$~\cite{nizkaya2014gas} are also included in Fig.~\ref{fig_beff2}(b), and we see that they are in a very good agreement with $b_{\mathrm{eff}}$ obtained with $\theta= \pi/6$ in a very large range of parameters. Finally, to examine the significance of $\theta$ more closely, in Fig. \ref{fig_beff2}(c) we plot $b_{\mathrm{eff}}$ against $\theta$. The calculations
are made using $\mu = 0.2$ and several area fractions of lubricant. We see that the effective slip is linear in $\theta$, which is, of course, a consequence of our first-order perturbation theory. It is also seen that the value of $b_{\mathrm{eff}}$ and a slope of these lines decrease with $\phi$, as could be expected from above results. This plot also confirms that the effect of $\theta$ on the effective slip length is very little compared to that of $\phi$ and $\mu$.

\section{Concluding remarks}
\label{concl}

%\textbf{1. Brief summary of results and why they are interesting. }

By means of a perturbation theory we have calculated the (normalized by $L$) effective longitudinal slip length, $b_{\mathrm{eff}}$, of a lubricant-infused surface, assuming that the meniscus protrusion angle, $\theta$, is small and that the viscosity of a lubricant is smaller or equal to that of liquid. Our theory provides considerable insight into slippage generated at lubricant-infused surfaces depending on the area fraction of lubricant, $\phi$, viscosity contrast, $\mu$, and protrusion angle, $\theta$. We have shown that the value of $b_{\mathrm{eff}}$ depends strongly on the viscosity contrast of two phases. In the limit of vanishing lubricant viscosity, $\mu = 0$, we recover results by~\citet{sbragaglia.m:2007}. In this case, where the correction to the effective slip length of a flat interface, $\theta b_{\mathrm{eff}}^1$, is largest, we have proposed a simple analytical formula, Eq.(\ref{ab3}), describing it accurately in a very large range of $\phi$, which probably includes its all experimentally relevant values. In the opposite case of equal viscosities of liquid and lubricant, $\mu = 1$, we have shown that finite $\theta$ does not influence the solution obtained by~\citet{wang2003} for filled with liquid grooves. Our work clarifies that in a very large range of $\mu$ and $\phi$, the correction $\theta b_{\mathrm{eff}}^1$ can be neglected, and $b_{\mathrm{eff}}$ can be accurately calculated from Eq.(\ref{eq:belyaev}) by \citet{belyaev.av:2010a} with a (scaled with $L$) local slip length at the lubricant area determined solely by $\phi$ and $\mu$.

%We expect that many of our results will have validity beyond this particular model whilst others are certainly specific to it.

%\textbf{2. How the results can be used or checked in experiment. How can they be extended? }

Our strategy can be extended to calculations of effective slip lengths for a liquid flow transverse to lubricant-infused stripes. \citet{lauga2009} have found the transverse effective slip length of the SH surface in the limit of $\phi \ll 1$ without restriction on the protrusion angle, $\theta$. Recent work~\cite{crowdy.d:2017} has concluded that leading order corrections to transverse and longitudinal effective slip lengths of SH grooves with $\theta \ll 1$, are identical. We are unaware of any previous theoretical work that has attempted to calculate transverse  $b_{\mathrm{eff}}$ for grooves with weakly protruding menisci and finite $\mu$, and the extension of our approach to this case would appear to be very timely.

Finally, we mention that our approach can be applied to compute slip lengths of grooves filled by a lubricant of higher viscosity than that of a liquid. Such LI surfaces can also reduce viscous drag~\cite{solomon.br:2014,keiser.a:2017}, but only slightly, since a viscous dissipation in a lubricant becomes significant. By combining perturbation approach with the reciprocity ideas \citet{crowdy.d:2017b} (see his supplementary material) calculated $b_{\mathrm{eff}}$ in the limit $\mu \gg 1$. It seems to be appropriate to calculate $b_{\mathrm{eff}}$ in a whole range of $\mu\geq 1$  by means of an approach used here.  

%Turbulent drag reduction?~\cite{fu.mk:2017}

%The range of validity of this asymptotic slip length profile is discussed in Appendix~\ref{sec_app_validity} for the case of rectangular grooves.

\begin{acknowledgments}
This research was partly supported by the Russian
Foundation for Basic Research (grant No. 18-01-00729).
\end{acknowledgments}

\appendix
\section{Boundary conditions at the lubricant-liquid interface}\label{ap:A}

Here we transform conditions of continuity of velocity and shear stress at the lubricant-liquid interface, Eq.(\ref{shstr}), to conditions imposed at $z=0$, where the effective slip is defined.

For small meniscus curvature the zero-order liquid velocity at $z=\theta \eta$ can be expanded about $U+u^{0}$, and to first order in $\theta$
\begin{eqnarray}
\left. \left(U+u^{0}\right)\right\vert _{z=\theta \eta }&\simeq&\left.  U+u^{0}+\theta \eta \partial _{z}\left(
U+u^{0}\right)  \right\vert _{z=0} \notag \\
&=&
\left.  u^{0}+\theta \eta \left(
1+\partial _{z}u^{0}\right)  \right\vert _{z=0} \label{uexp1}
\end{eqnarray}%
If we extrapolate the zero-order lubricant velocity to $z=\theta \eta$, an expansion about $u^{0}_l$ gives
\begin{equation}
\left. u^{0}_l\right\vert _{z=\theta \eta }\simeq \left. \left( u^{0}_l+\theta \eta \partial _{z}
u^{0}_l \right) \right\vert _{z=0}  \label{uexp3}
\end{equation}
The first-order velocities can be expanded as
\begin{equation}
\left. \theta u^1\right\vert _{z=\theta \eta }\simeq \left. \theta u^1\right\vert _{z=0}, \left. \theta u^1_l\right\vert _{z=\theta \eta }\simeq \left. \theta u^1_l\right\vert _{z=0}  \label{uexp2}
\end{equation}%
Since the velocities of liquid and lubricant at $z=\theta \eta$ are equal, from Eqs.(\ref{exp}) and  (\ref{vt0}) it follows that at $z=0$
\begin{equation}
u^{1}+\eta \left(
1+\partial _{z}u^{0}\right) \simeq u^{1}_l+\eta \partial _{z}u^{0}_l,  \label{sb3}
\end{equation}%
Using Eq.(\ref{sh0}) this equation can be reduced to Eq.(\ref{sb5}).

Similarly, we construct expansions for shear stresses, and using  Eqs.(\ref{uexp1}-\ref{uexp2}) derive
\begin{eqnarray}%
\left. \left( \mathbf{n\cdot \nabla }\right) u \right\vert _{z=\theta \eta }
\simeq   1 &+&
\partial _{z}u^{0} +  \\
&& \left. \theta \left( \eta \partial
_{zz}^{2}u^{0}-\partial _{y}\eta \partial _{y}u^{0}+\partial _{z}u^{1}\right)\right\vert _{z=0}\nonumber ,
\label{sh3}
\end{eqnarray}%
\begin{eqnarray}%
\left. \left( \mathbf{n\cdot \nabla }\right) u_l \right\vert _{z=\theta \eta }
\simeq  && \partial _{z}u^{0}_l + \\
&&\left. \theta \left( \eta \partial
_{zz}^{2}u^{0}_l-\partial _{y}\eta \partial _{y}u^{0}_l+\partial _{z}u^{1}_l\right)\right\vert _{z=0} \nonumber
\label{sh4}
\end{eqnarray}%
Since $\partial _{zz}^{2}u^{0}= -\partial _{yy}^{2}u^{0}$ and $\partial _{zz}^{2}u^{0}_l= -\partial _{yy}^{2}u^{0}_l$, and using
Eq.(\ref{sh0}) we obtain the condition for a tangential stress given by Eq.(\ref{sb1}).%

\section{Approximate formulas for $b^{1}_{\mathrm{eff}}$ at $\mu = 0$}
\label{ap:B}

In this Appendix, we derive a simple formula for $b_{\mathrm{eff}}^{1}$ and discuss
its asymptotics.
It has earlier been shown that when $\phi \ll 1$, the velocity profile, Eq.(\ref{u0}), is close to elliptic~\cite{philip.jr:1972}:%
\begin{eqnarray}
u^{P}(y,0) &\simeq &u_{\max }^{P}\left( 1-\frac{4y^{2}}{\phi ^{2}}\right)
^{1/2},  \label{ab1} \\
\partial _{y}u^{P}(y,0) &\simeq &-\frac{4u_{\max }^{P}y}{\phi ^{2}}\left( 1-%
\frac{4y^{2}}{\phi ^{2}}\right) ^{-1/2},
\end{eqnarray}%
where $u_{\max }^{P}$ is given by Eq.(\ref{upm}).
However, our calculations (see symbols in Fig.~\ref{fig_deltabeff}(b)) suggest that this conclusion remains valid for much larger $\phi $. We can then calculate the first-order correction to the slip length given by Eq.(\ref{sb4}) as
\begin{equation}
b_{\mathrm{eff}}^{1} \simeq \frac{8\left( u_{\max }^{P}\right) ^{2}}{\phi ^{3}}%
\int_{0}^{\phi /2}y^{2}dy = \frac{\left( u_{\max }^{P}\right) ^{2}}{3}
\label{ab3_1}
\end{equation}
Therefore,
\begin{equation}
b_{\mathrm{eff}}^{1} \simeq  \frac{\Arch^{2}\left[ \sec \left( \frac{\pi \phi }{2}\right) \right] }{3\pi ^{2}}
\label{ab3}
\end{equation}
\\
\begin{figure}[h!]
\centering
\includegraphics[width=\columnwidth]{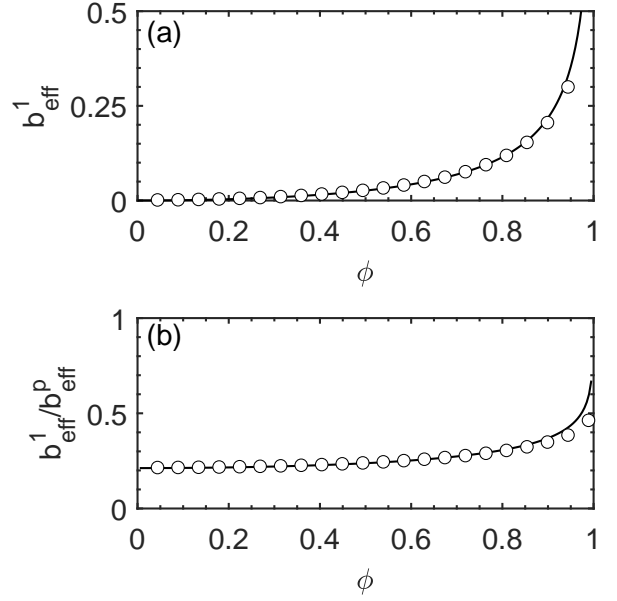}
\caption{(a) The first-order slip length, $b_{\mathrm{eff}}^{1}$,  for the case of $\mu = 0$ plotted against $\phi$. Solid curve is calculated from Eq.(\ref{ab3}), symbols show $b_{\mathrm{eff}}^{1}$ given by Eq.(\ref{sb4}). (b) The same $b_{\mathrm{eff}}^{1}$, normalized by $b_{\mathrm{eff}}^{P}$.}
\label{Fig_ap2}
\end{figure}

Fig.~\ref{Fig_ap2} includes predictions of Eq.(\ref{sb4}) along with a theoretical curve calculated from Eq.(\ref{ab3}). The ratio $b_{\mathrm{eff}}^{1}/b_{\mathrm{eff}}^{P}$ is also plotted. We note that the fit is extremely good for $\phi \leq 0.9$, but there is a discrepancy at larger $\phi$. We see that when solid fraction is getting very small, Eq.(\ref{ab3}) overestimates $b_{\mathrm{eff}}^{1}$. Indeed, in the limit $1-\phi \rightarrow 0$, Eq.(\ref{ab3}) reduces to
\begin{equation}
b_{\mathrm{eff}}^{1}\simeq \frac{2}{3\pi ^{2}}\ln^{2}\left( 1-\phi \right),
\label{ab4}
\end{equation}
and $b_{\mathrm{eff}}^{1}/b_{\mathrm{eff}}^{P}\propto - \ln \left( 1-\phi \right)$, i.e. diverges logarithmically. We recall that \citet{sbragaglia.m:2007} predicted $b_{\mathrm{eff}}^{1} \propto - \ln \left( 1-\phi \right)$, which implies that $b_{\mathrm{eff}}^{1}/b_{\mathrm{eff}}^{P}$ is always finite. We can, therefore, conclude that Eq.(\ref{ab3}) cannot be employed at a very large $\phi$, and this, of course, indicates that velocity profiles are no longer elliptic.

We finally note, that when $\phi\rightarrow 0$,

\begin{equation}
b_{\mathrm{eff}}^{1}\simeq \frac{\phi^2}{12},
\label{ab5}
\end{equation}
 and $b_{\mathrm{eff}}^{1}/b_{\mathrm{eff}}^{P} \simeq 2/3\pi (1+\pi^2 \phi^2/24)$. These, obtained in the low $\phi$ limit, formulas are surprisingly accurate up to $\phi \simeq 0.5$, i.e. have validity well beyond the range of their formal applicability.

\bibliographystyle{rsc}
\bibliography{meniscusbib,viscosity}

\end{document}